\documentclass[conference]{IEEEtran}
\usepackage{blindtext, graphicx}
\usepackage{todonotes}
\usepackage{framed}
\usepackage{multirow}
\usepackage{url}

\usepackage[inline]{enumitem}

\ifCLASSINFOpdf
\else
\fi

\begin{document}
\title{On the Presence of Green and Sustainable Software Engineering in Higher Education Curricula}

\author{\IEEEauthorblockN{Damiano Torre}
\IEEEauthorblockA{SQUALL, Carleton University, Canada\\
Alarcos, UCLM, Spain\\
dctorre@sce.carleton.ca}
\and
\IEEEauthorblockN{Giuseppe Procaccianti}
\IEEEauthorblockA{Software and Services\\
Vrije Universiteit Amsterdam \\The Netherlands\\
g.procaccianti@vu.nl}
\and
\IEEEauthorblockN{Davide Fucci}
\IEEEauthorblockA{M3S\\
University of Oulu, Finland\\
davide.fucci@oulu.fi}
\and
\IEEEauthorblockN{Sonja Lutovac}
\IEEEauthorblockA{Faculty of Education\\
Department of Educational Sciences and Teacher Education\\
University of Oulu, Finland\\
sonja.lutovac@oulu.fi}
\and
\IEEEauthorblockN{Giuseppe Scanniello}
\IEEEauthorblockA{DiMIE\\
University of Basilicata, Italy\\
giuseppe.scanniello@unibas.it}
}

\maketitle
\begin{abstract}
%\todo[author=Damiano,inline,color=pink]{Update it}
Nowadays, software is pervasive in our everyday lives. 
Its sustainability and environmental impact have become major factors to be considered in the development of software systems. 
Millennials--the newer generation of university students--are particularly keen to learn about and contribute to a more sustainable and green society. 
The need for training on green and sustainable topics in software engineering has been reflected in a number of recent studies.
The goal of this paper is to get a first understanding of what is the current state of teaching sustainability in the software engineering community, what are the motivations behind the current state of teaching, and what can be done to improve it.
To this end, we report the findings from a targeted survey of 33 academics on the presence of green and sustainable software engineering in higher education. 
The major findings from the collected data suggest that sustainability is under-represented in the curricula, while the current focus of teaching is on energy efficiency delivered through a fact-based approach.
The reasons vary from lack of awareness, teaching material and suitable technologies, to the high effort required to teach sustainability.
Finally, we provide recommendations for educators willing to teach sustainability in software engineering that can help to suit millennial students needs.
\end{abstract}

\IEEEpeerreviewmaketitle

\section{Introduction}\label{sec:introduction}
The latest measurement (December 2016) of the global concentration of CO2 in the atmosphere, i.e. the primary driver of contemporary climate change, has reached 405.25 parts per million (ppm), the highest in recorded history.\footnote{http://climate.nasa.gov/vital-signs/carbon-dioxide/}  
Serious actions must be taken to avoid hitting the ``point of no return,'' which is when no amount of cutbacks on emissions will save us from the potentially catastrophic repercussions of global warming. 
%It is not surprising that several initiatives, trying to contribute to a more sustainable society, are gaining importance worldwide across vastly different fields.
Millennials--the newer generation of university students--are particularly keen to learn about  and contribute to a more sustainable and green society~\cite{6227999}. 
However, a survey of 3860 software engineering practitioners working at IBM, Google, ABB, and Microsoft, showed that the current higher education curriculum does not prepare them to tackle sustainability, although these practitioners were willing to learn about sustainability~\cite{manotas2016empirical}.
For example, one of the respondent, referring to the general lack of concern towards rising sustainability issues, commented: \textit{``I would love to have more education [...] for designing and investigating battery lifetime! Anything to help raise awareness and break through attitude barriers''}.
%In this paper we focus on green and sustainability in the area of software engineering(GSSE), and specifically in the context of higher education curricula.
% Software development has become much more widespread, as it is pervasive in our daily lives.
These issues should be considered from the prospective of millennials, as they are not only the main consumers, but also the main producers of software, as software development is taught in schools, from primary to higher education~\cite{gutbrod2012software}.
% Another preliminary work [TR Dam-Dav] shows that after searching for green and sustainable topics in the worldwide top 10 universities websites\footnote{http://www.topuniversities.com/university-rankings/world-university-rankings/2015}, these topics were still not present in the curricula of those universities. 
%The results of constitute the main motivation for the study reported in this paper.
% \todo[author=Giuseppe, inline, color=yellow]{A bit weak: we only talk about energy/battery efficiency, we should have a more substantial motivation about sustainability}
% \todo[author=Davide,inline,color=green]{Basing the motivation on a TR seems weak in hindsight. Move it to Section II and relate it directly to [22]}
% \todo[author=Damiano, inline, color=pink]{che dite?}

Starting from the results of the study by Manotas et al.~\cite{manotas2016empirical}, and the far-reaching importance of topics such as greenability and sustainability~\cite{calero2015green} especially for millennial teaching and learning ~\cite{6227999}, we surveyed 33 researchers with experience in green and sustainability software engineering ~\cite{Becker} (GSSE) to appraise this field, and to offer a first support to the software engineering (SE) community in the development of courses and curricula addressing GSSE.
%\todo[author=Giuseppe, inline, color=yellow]{@Damiano, How many did we have in the end?}
This study has the merit to be the first that surveys the state of education as well as some of  the factors necessary for the development of a higher-education curricula in GSSE, which might be of great value for millennials.
%\todo[author=Davide,inline,color=green]{Too strong claim?}
%\todo[author=Giuseppe,inline,color=yellow]{Maybe. Ci sono dei surveys, ma non specifici per SE. Inoltre non so quanto si possano definire "empirical" investigations. Io lo terrei}
In this paper, we make the following contributions: \textit{(i)}~a survey of 33 experts in  GSSE to quantitatively assess themes related to GSSE education; \textit{(ii)}~a contextualization of the state of GSSE education, and suggestions for the development of related curricula; and \textit{(iii)}~recommendations for educators in GSSE.

The rest of this paper is organized as follows: In Section \ref{sec:background}, we report the most salient studies investigating the GSSE curricula. In Section \ref{sec:method}, we describe the research methodology used in our study. %, i.e., survey, used to conduct this study. 
In Section \ref{sec:results}, we highlight our results, whereas we present our recommendations in Section \ref{sec:discussion}. In Section~\ref{sec:threats}, we discuss on possible threats that could affect the validity of the observed results. Final remarks and future work conclude the paper in Section~\ref{sec:conclusion}. 

\section{Related Work}\label{sec:background}
GSSE is an increasing priority being it in the spotlight for both professionals and academic researchers~\cite{Lago2013-iz, Lago2014-ir}.
However, there is a lack of evidence and consolidated knowledge about the topic~\cite{Lago2014-ai, Pang2015-el}.
This represents a risk for educators that want to introduce sustainability in SE programs and curricula.
Nonetheless, sustainability and energy efficiency are regarded as key competences for future software engineers and developers \cite{Lago2014-ai, Lago2015-zq, Penzenstadler2011-wy, Pang2015-el}.

A comprehensive list of courses related to IT and IS sustainability can be found in a survey by England et al. \cite{England2012-ir}. 
The survey concludes that \textit{``very few institutions of higher education were found to have Green IT/IS degree programs''}.
On the one hand, most of the educational offers in the field are actually provided as training or certification activities for professionals, rather than proper academic courses taught at Bachelor or Master level.

A survey by Merkus~\cite{Merkus2013-km} identified initiatives in a total of 19 universities, 10 in Europe and 9 outside Europe. 
Most consist in individual modules that either focus on Green IT specific subjects, or address Green IT related topics within a technical subject (e.g. sustainable business processes within a module on virtual organizations). 
In particular, Leeds University (UK) offers a full Master program on Sustainable Computing, and the University of Lorraine (France) offers a module related to Green Software.

The observatory for Engineering Education for Sustainable Development (EESD) identified the following challenges for introducing sustainability into university curricula~\cite{Motrel2008-fc}:
\begin{enumerate}
	\item Defining appropriate and relevant content for an engineering education;
	\item Providing inspiration through sustainability activities within and through the campus;
	\item Increasing research activities as a support for education;
	\item Gaining acceptance for the importance of sustainability within the university leadership.
\end{enumerate}
In the report focused on technical universities, the EESD observatory also states that only one institute, out of the 55 that participated worldwide, could be considered as an ``inspiration'' for sustainability activities.

A few works exist that propose frameworks and guidelines on how to deliver content on sustainability in traditional education. 
For example, Mann et al.~\cite{Mann2010-zq} provide a framework meant for educators to design modules/programs addressing sustainability, that classifies sustainability-focused education approaches in three types: centralized (i.e. concentrate sustainability topics in one or two focused courses), distributed (i.e. address sustainability topics across all the courses in the curriculum), and blended (i.e. a mix of the the previous approaches). 
Sammalisto et al. \cite{Sammalisto2007-tn} performed a study on the integration of sustainability in higher education, classifying courses across different sustainability dimensions. 
The concept of multi-dimensional sustainability is quite well-established in the literature \cite{Bruntland1987-kz, Goodland2002-so, lago-cacm, Razavian2014-mn}. In our paper, we also use this concept, by explicitly considering four dimensions: environmental, economic, social and technical.
The study by Sammalisto et al. \cite{Sammalisto2007-tn} concludes that a proper feedback system has to be in place between educators and university administrators to demonstrate the value and importance of the integration of sustainability. 

In a first preliminary analysis of the top 10 universities curricula~\cite{Torre2016}, we showed that, although energy and sustainability related issues are topics of interest for few engineering courses, none explicitly addresses \textit{Green Software Engineering}, nor \textit{Sustainable Software Engineering}.

On the other hand, the topic of educating millennials is starting to be addressed only recently. The higher education pedagogy literature suggests that teaching should be adapted to the characteristics of this particular segment, as this generation of students are ``visually focused,'' and accustomed to hyper-personalized experiences \cite{6227999}. They also value \textit{doing} more than \textit{knowing}, and--as they are used to deal with a large amount of choices--embrace multitasking \cite{mcglynn2005teaching}.

\section{Methodology}\label{sec:method}
In this study, we perform a survey to investigate the state of higher education in GSSE. The survey methodology is a well-established technique for collecting data about features, behavior, or opinions of a specific group of people, representative of a target population~\cite{pinsonneault1993survey}. 
Specifically for this study, we chose to use the \textit{on-line survey} method, as it allowed us to obtain information from a relatively large number of experts in a short amount of time. Besides data collection, online surveys also simplify data categorization and analysis.
%This section details the survey goal, design, execution, and pre-analysis data validation.

\subsection{Goal and Research Questions}
The United Nation defines the Education for Sustainability practice as a \textit{learning process aimed at equipping students, teachers and institution with the knowledge needed to achieve economic prosperity while restoring the health of the living systems upon which our lives depend on}.\footnote{\url{http://www.unesco.org/new/en/education/themes/leading-the-international-agenda/education-for-sustainable-development/education-for-sustainable-development/}}

The main goal of this survey is to assess the reasons for the current state of teaching GSSE in higher-education, as well as identifying challenges and recommendations.
To do that, we tackled the following \textit{Research Questions (RQ):}
\begin{description}
\item[\textbf{RQ1:}] What is the background of academics investigating in the area of GSSE?
\item[\textbf{RQ2:}] How are the GSSE topics considered in the academic community? %to be improved
\item[\textbf{RQ3:}] How involved are the academics investigating in the area of GSSE in teaching GSSE topics in higher-education?
\item[\textbf{RQ4:}] What are the challenges and recommendations seen by academics investigating in the area of GSSE in teaching GSSE topics in higher-education?
\end{description}

\subsection{Survey Design}
The survey is designed to capture the information needed to answer our research questions.
We made sure that the questions are relevant to the context of GSSE topics in higher education.  
Our design follows the software engineering survey guidelines \cite{kitchenham2008personal,linaker2015guidelines,punter2003conducting}. 
%\todo[author=Davide,inline,color=green]{These are guidelines about surveys to understand software engineering practices. Our survey is related to education. Just a thought}

\subsubsection{Identifying Target Audience}
The first step to conduct a survey consists in defining a target population.
In our study, the target population is composed by researchers that, at the time of writing, are or have been involved in the organization of workshops on GSSE and/or published papers in these workshops. 
To collect the information about the target population we looked at the proceedings of the following workshops:
\begin{itemize}
	\item Requirements Engineering for Sustainable Systems\footnote{http://web.csulb.edu/~bpenzens/re4susy} (RE4SuSy);
	\item Green and Sustainable Software\footnote{http://greens.cs.vu.nl/} (GREENS);
	\item Green and Sustainable Software Systems\footnote{http://alarcos.esi.uclm.es/eseiw2016/megsus/home} (MeGSuS);
	\item Green in Software Engineering\footnote{http://alarcos.esi.uclm.es/ginseng2016} (GInSEng). 
\end{itemize}
We stored the data (name, last name, email, and affiliation) regarding the target population into a Google spreadsheet. 
Since we recruited subjects from workshops' proceedings on GSSE, our approach to sampling was non-probabilistic~\cite{kitchenham2008personal}. 
Our population consisted of 165 academics.
\subsubsection{Survey Questions}
Our survey instrument was a questionnaire \cite{linaker2015guidelines}. 
In order to develop a survey that would adequately gather the information needed to answer our RQs, we developed a questionnaire of four sections with a total of 18 survey questions (SQ). %, as shown in Table~\ref{tbl:survey}. 
The first part of the questionnaire contained three questions gathering the general background data of the respondents (see Section \ref{sec:qBackground}). 
The second section contains two questions probing the importance of GSSE topics (see Section \ref{sec:qGSSE}).
Ten questions---specified in the third section of the questionnaire---focused on education and teaching with respect to GSSE.
In this section we investigate specific details of courses in the area of GSSE (see Section \ref{sec:qEducation}). 
Finally, the last section includes three questions prompting respondents for new ideas and future challenges for GSSE courses (see Section~\ref{sec:qIdea}).

\subsubsection{Survey Execution}
We collected data through an on-line questionnaire created by means of a web-based questionnaires tool.\footnote{Google Forms - http://www.google.com/forms} 
%In general, it has been observed that web-based questionnaires guarantee high return rates~\cite{jedlitschka2007relevant}. 
The survey was conducted between 20th September 2016 and October 20th.
The URL of the survey has been emailed directly to the selected target audience.

\subsubsection{Pre-analysis Considerations and Data Validation}
We collected 33 questionnaires correctly filled in, thus obtaining 20\% as response rate. 
%Detailed results of the survey are discussed in Section~\ref{sec:results}. 
Given the  size of our sample, we did not perform hypothesis testing nor extract dependency statistics (e.g., correlations) from the data. Using quantitative analysis on our data could have resulted in an overstatement of effects that could lead to misinterpretation of our results \cite{Maxwell01072010}.

\section{Survey Results}\label{sec:results}
In this section, we report the results for each of the four questionnaire sections.
%A summary of the results for research questions SQ1-SQ4, SQ6-SQ8, SQ10-SQ15 and SQ17 are presented in Table \ref{tbl:results}. 
%A summary of the results for research questions are provided in the following sub-sections.
\subsection{Background Information}\label{sec:qBackground}
In this questionnaire section, we aim at collecting general background information on the respondents, to perform a demographic analysis needed to answer \textit{RQ1}. This section contains close-ended questions on nationality (SQ1), age (SQ2), and seniority (SQ3). 

\subsubsection{SQ1. What is your nationality?}
%The first survey question asked respondents about their location. 
The final dataset had 33 valid respondents from 13 different countries. The countries in which the respondents work are very varied: Spain (5), Germany (5), Portugal (4), Italy (3), France (3), China (3), United Kingdom (2), Brazil (2), United States (2), India (1), Colombia (1), Canada (1), and Belgium (1).

%\todo[author=Davide,inline,color=green]{Perche chiediamo la professione ed eta, e poi chiediamo di nuovo la professione nella prossima domanda? - Sistemato}
\subsubsection{SQ2. What is your age range?}
The age ranges of respondents were: 8 respondents with less than 30 years, 17 between 30 and 45 years, and 8 had more than 45 years. 

\subsubsection{SQ3. What is your current position?}
The majority of those taking part have a professor position (19, 57.7\%): 9 Assistant Professor, 5 Associate Professor and 5 Full Professor. 
The rest of respondents (14, 42.3\%) were: 9 PhD student, 3 Post-doc, and 2 \textit{other}. 
According to the sample covered by this survey, the results allows us to state that---considering Assistant Professor, Post-doc and PhD student together (21, 63.7\%) and the fact that the majority of respondents have less than 45 years (25, 75.7\%)---the area of GSSE is mostly the focus of a young generation of academics.

\subsection{Green and Sustainable Software Engineering}\label{sec:qGSSE}
In this questionnaire section, we answer \textit{RQ2} by asking respondents about their prioritization of the sustainability dimensions (SQ4) and importance of GSSE topics in their institution (SQ5).

\subsubsection{SQ4. How would you rank the following dimensions: Economic, Environmental, Social, and Technical?}
According to their experience and positionality regarding to GSSE topics, we asked the respondents about how they would rank, on a four-point Likert scale (from \textit{Very Important} to \textit{Not Important}), the following four dimensions of software sustainability:
\begin{description}
\item[Economic:] Very Important (8, 24.2\%), Moderately Important (15, 45.5\%), Slightly Important (9, 27.3\%), Not Important (1, 3\%);
\item[Environmental:] Very Important (22, 67.7\%), Moderately Important (6, 18.2\%), Slightly Important (5, 15.2\%), Not Important (0);
\item[Social:] Very Important (18, 54.5\%), Moderately Important (10, 30.3\%), Slightly Important (2, 6.1\%), Not Important (3, 9.1\%);
\item[Technical:] Very Important (22, 66.7\%), Moderately Important (10, 30.3\%), Slightly Important (1, 3\%), Not Important (0).
\end{description}

The results suggest that all the four dimensions used in the area of GSSE are considered very important. Indeed, the economic dimension is considered moderately important.
%Figure~\ref{fig:bubble} shows the frequencies of combining the results from the different Likert items of SQ5 in a bubble chart. A bubble plot is basically two x–y scatter plots with bubbles in category intersections. This synthesis method is useful to provide a map and it gives a quick overview of a research field \cite{Petersen:2008:SMS:2227115.2227123}.  
%Instead of the bubbles, we have pie charts detailing other results---i.e., the bubbles describe the age ranges for each answer presented in SQ2. 
%For example, out of the 15 respondents identifying the Economic dimension as moderately important, 4 had less than 30 years, 8 between 30 and 45 years, and 3 more than 45 years.
%\begin{figure}
 %  \centering
  %\includegraphics[width=.49\textwidth]{figures/Bubble_C_Green}
 %\caption{Participants ranks of the Green and Sustainable dimensions}
%\label{fig:bubble}
%\end{figure}

\subsubsection{SQ5. How important is the topic of Green and Sustainable Software in the Software Engineering curricula of your university?}
%Following the bove question SQ5, this time participants were asked about how important is the topic of GSSE in the software engineering curricula of their universities. 
The gathered answers suggested that GSSE topics are not important or partially important. 
The answers in in detail are:
\begin{enumerate*}
\item \textit {Not important} (20, 60.6\%): Green and Sustainable Software is not taught in the SE curricula of my university;
\item \textit {Partially important} (10, 30.3\%): Green and Sustainable Software is taught in a semester-long or shorter course;
\item \textit {Very important} (3, 9.1\%): Green and Sustainable Software is an important topic taught in more than one semester-long course.
\end{enumerate*}

%From these results we can highlight the challenge and frustration of the majority of respondents (60.6\%).
%On one side they are doing research on GSSE, on the other side they see their universities not focusing on this area. 

\subsection{Education in Green and Sustainable Software}\label{sec:qEducation}
In order to answer \textit{RQ3}, we ask the respondents about their experience in teaching GSSE topics. We ask if they participate to academic events (SQ6) focused on GSSE, if they teach any course related to GSSE (SQ7), and if not, if they teach any GSSE topics in traditional SE courses (SQ8). The respondents, who answered `yes' to these questions, were successively asked to provide some additional information on their courses: name (SQ8), number of students (SQ10), duration (SQ11), evaluation method (SQ12), third--party involvement (SQ13) required effort (SQ14) and specific focus (SQ15).

\subsubsection{SQ6. Do you regularly participate in academic events related to software sustainability? If so, which ones?}
The respondents are involved in following academic events related to GSSE: GREENS (6 respondents), WSSPE\footnote{http://wssspe.researchcomputing.org.uk/} (2 respondents), RE4SuSy (2 respondents), SEIS\footnote{http://2016.icse.cs.txstate.edu/seis} (2 respondents), and MeGSuS, ICT4S\footnote{http://ict4s.org/}, GInSEng and ESEIW\footnote{http://alarcos.esi.uclm.es/eseiw2016/} (1 respondents). The answers such as \textit{``I give lecture or keynote''}, \textit{``I attend conference''}, were not reported because not enough specific to concrete events.

\subsubsection{SQ7. Do you currently (or ever) teach any course related to software sustainability?}
60.6\% of the respondents (21 out of 33) did not ever teach any course related to GSSE topics despite their expertise.

\subsubsection{SQ8. As you do not teach any course focused on sustainable software, do you teach sustainability topics in your traditional software engineering courses? (e.g. in a programming course, best practices for energy-efficient code). If you replied ``no'', why not?}
The 13 out of 21 respondents that did not ever teach a course on GSSE 
%were asked if they teach sustainability topics in their traditional software engineering courses. 
%The 81\% (17 of 21 respondents considered) 
replied that they do not teach GSSE topics in their SE course for the following reasons:
\begin{itemize}
   \item  \textit{``GSSE topics are not part of the curricula of their university,''} (two respondents).
   \item  \textit{``there is not time to explore such issues within the duration of the SE course.''} (six respondents).
   \item \textit{``they are not teaching right now but if they have the opportunity they would like to introduce these topics''} (two respondents).
   \item \textit{``I prefer to follow the classical curriculum and focusing on technical aspects only''} (one respondent).
   \item \textit{``I am teaching databases and has no feedback on energy-efficient data modelling or query design''} (one respondent).
   \item \textit{``I am teaching programming languages and algorithms for freshman students and that they don't have the maturity for dealing with this topic''} (one respondent).
\end{itemize}
\subsubsection{SQ9. What is the name of the course?}
We asked the respondents who currently teach courses related to software sustainability, and to other which teach sustainability topics in their traditional software engineering courses to name those courses. 
The courses that seems more related to sustainability and green topics were: 
\begin{enumerate*}
\item Sustainable Software Engineering, 
\item Requirements Engineering for Sustainability,
\item Green Software, 
\item Special Topics: Developing Energy Efficient Applications, 
\item Service oriented design, 
\item Green Lab, 
\item Digital service eco-design 
\item Energy-efficient programming
\end{enumerate*} 
Other traditional SE courses where the respondents said to include GSSE topics were: 
\begin{enumerate*}
\item Introduction to Programming Systems Design,
\item Requirements Engineering, 
\item Software Architectures, 
\item Analysis and Testing for Software,
\item Parallel Computer Architectures, clusters and grids, 
\item Machine Learning Applied, 
\item Advanced tools for software engineering, 
\item Software Analysis and Transformation, and 
\item Software Engineering.
\end{enumerate*}

\subsubsection{SQ10. How many students do you have on average?}
This question looks at how many students attend courses that included GSSE topics. 
The 50\% of respondents said that their GSSE courses (or traditional SE course with some GSSE topics) have less than 20 students, 37.5\% between 20 and 50, 6.3\% between 50 and 100, and 6.3\% more than 100.

\subsubsection{SQ11. What is the duration of the course?}
We asked the respondents the duration of their GSSE courses (or traditional SE course with some GSSE topics).
The 75\% of respondents stated that the courses are one semester long, 18.8\% said that they are less than one semester, and the 6.3\% said that are two semesters long.

\subsubsection{SQ12. Is the course exam-based or assignment-based?}
This question inquired about if the courses are exam-based or assignment-based.
The 43.8\% of the respondents said that the courses are assignment-based, the 12.5\% are exam-based, and 43.8\% of them said to follow other teaching procedures.

\subsubsection{SQ13. Do you involve third parties in your course? (e.g. public/private stakeholders). If so, please describe what type of stakeholders and their involvement.}
The 81.3\% of respondents said they do not involve any third parties in their course; the 18.8\% instead said to use external stakeholders according to their expertise needs---i.e., PhD students giving seminars in their area of research, and IT software industry professionals.

\subsubsection{SQ14. Compared with other traditional courses, how difficult was to design the course(s) on Green and Sustainable Software topics? Please motivate your answer} 
When respondents were asked to compare with other traditional courses, how difficult is to design the course(s) on Green and Sustainable Software topics (from \textit{Very Difficult} to \textit{Not Difficult}), our result shows that 25\% considered this activity very difficult because: 
\begin{enumerate*}
\item it involves both software and hardware considerations;
\item it is mainly due to the novelty of the field;
\item there is no real organized information of material to teach;
\item there are already ways to accurately measure energy in software (in an easy and distribute manner);
\item there is not detailed knowledge of practices which can improve/deteriorate energy consumption in software for students to play around with;
\item the students are not aware about digital services environmental impacts.
\end{enumerate*}

The 53.6\% of respondents considered the activity moderately difficult because:
\begin{enumerate*}
\item the sustainability is very hard to realize without further information;
\item there is a wide selection of topics that could be related with GSSE and it is not possible to fit so much into one semester. 
\end{enumerate*}

The 6.3\% of the respondents said that this activity is slightly difficult because the GSSE topics are synergically included in their own research. 
Finally, we saw that there were not difficulties in designing GSSE courses for the 12.5\% of respondents.

%\todo[author=Davide,inline,color=green]{Sembra che diciamo il contrario di SQ14, o sbaglio? - Dam: si non ha senso, s14 e non fornisce spunti e ci serve spazio, la cancello}

%\todo[author=Davide,inline,color=green]{SQ13 dovrebbe essere SQ14}
%\subsubsection{SQ15. Have you faced any problems in designing the course(s) on Green and Sustainable Software topics? If so, which ones?}
%Even though the 53.6\% of respondents considered the activity of design a GSSE course moderately difficult in SQ13, when we asked if they have faced any problems in designing those course(s), the 87.5\% answered "No". 
%The other 12.5\% that faced some problems did not provide any explanation.

\subsubsection{SQ15. Which green and sustainable software aspect (s) were the focus of your course(s)? (e.g., Energy Efficiency, Maintainability, Social Sustainability, etc)}\label{sec:sq15}
In this question we asked which green and sustainable software aspect(s) were the focus of the course(s) taught by the respondents. 
From the answers to this question we ranked the topics by popularity:
\begin{enumerate*}
\item Energy efficiency (10 respondents);
\item Environmental and social sustainability (three respondents);
\item Maintainability (two respondents);
\item Architecture (one respondent);
\item Refactoring (one respondent);
\item Economical sustainability (one respondent);
\item Technical sustainability (one respondent).
\end{enumerate*}

\subsection{Ideas and challenges for the future}\label{sec:qIdea}
Finally in this section we answer \textit{RQ4} by identifying three open-ended questions.	Fist of all we ask respondents if there is a need for more courses realted to GSSE topics (SQ16). In order to understand the topics (SQ17) of possible GSSE courses, as well as the challenges (SQ18) they pose, we analyzed the answers using thematic analysis \cite{joffe20044}.
For each answers we labelled the text using open coding. 
Answers can have one or more labels.   
\subsubsection{SQ16. Do you think there is a need for more courses related to software sustainability?}
We asked this preliminary question to understand whether the respondents believe that the current state of teaching regarding sustainability is appropriate or not. 
The vast majority of the respondents (97\%) believe that there is the need for more courses related to sustainability in higher education curricula---i.e., they answered ``yes''.
Given the background of the respondents in the sample, such result was expected. 
However, it supports the motivation of this paper---i.e., the recognition that GSSE has not entered the university curricula, and therefore the need for improving the state of teaching sustainability topics. 

\subsubsection{SQ17. What type of software sustainability courses would you propose in a Software Engineering curricula?}\label{sec:q17}
The goal of this question is unfold ideas about the topics to be taught, and their characteristics with respect to sustainable software engineering.
We gathered 31 answers to this question, and extrapolated the following themes:
\begin{itemize}
	\item Topics: 14 respondents explicitly indicated the topics that should be addresses. Out of which, 11 indicated \textit{energy} and \textit{energy-efficiency} as the main topics to be taught, whereas three indicated \textit{performance} and \textit{performance optimization}. Three respondents believe that social aspects of sustainability should be taught in conjunction with technical topics. We identified six answers reporting only a broad indication of the topics (e.g., Green IT, or sustainable software engineering). Three respondents did not suggested specific topics, but rather believe that the they should be explicitly selected from the current research work. 
	\item Scope: 12 respondents gave indication of the scope that the teaching should span over. In particular, six respondents reported that GSSE topics should be taught for a specific aspect of software engineering (e.g., software architectures, refactoring). The remaining six indicated that the topics should cover a spectrum of aspect---i.e., all the software development and maintenance phases. 
	\item Organization: Five answers explicitly indicated that GSSE topics should be taught in a standalone course (one respondent envisioned a one-semester course). Two respondents, on the other hand, believe that GSSE topics should be distributed over existing courses on software engineering. 
\end{itemize}   
\subsubsection{SQ18. What challenges, if any, do you see in teaching software sustainability in higher education?}
The goal of this question was to prompt the respondents to identify the main challenges that would occur when establishing teaching modules (or courses) about GSSE.
We gathered 33 answers for this question, where four respondents reported no challenges in teaching GSSE, on the other hand from the other 29 ones, we extrapolated the following themes:
\begin{itemize}
	\item \textit{Awareness}: The main challenge (ten answers) is considered to be the lack of awareness regarding GSSE. Three respondents explicitly indicated the students as not being aware of sustainability as an issue for software engineering, whereas other three indicated the institutions and its goals to not be concerned with sustainability. One respondent indicated the lack of awareness from industry---as future employer of the students---as the main issue. The remaining answers generically reported awareness as an issue.
	\item \textit{Teaching material}: The other main challenge (six answers) appears to be the lack of teaching material. In particular, the respondents believe that there is not enough mature lecturing, project work, and assignment material that could be used in the classroom.
	\item \textit{Effort}: In six answers we found the large effort required to implement a course or module in GSSE to be the main challenge. The respondents argument that such effort is due to the novelty and multidisciplinary nature of GSSE. Two of the respondents added that more effort is required also from the students who would need to master other competences before being taught GSSE.
	\item \textit{Technology}: The lack of tools and technologies that can be used for didactic purposes is a theme that emerged in five answers. Two respondents explicitly mention the lack of access to specific hardware as a challenge to teach GSSE.
\end{itemize}

%\todo[author=Davide,inline,color=green]{Questa frase sembra un po appesa - Dam: apposto, spostata sopra}

%\todo[author=Davide,inline,color=green]{Dobbiamo cercare di azzecarci il discorso sui millennials} 
\section{Recommendations}\label{sec:discussion}

In this section, we provide three recommendations for educators based on the results of the survey, drawing from the current higher education research.

It is well established that pre-conceptions or beliefs educators have about teaching will have an impact on their teaching practices \cite{kember1997reconceptualisation,richardson1996role}. 
If academics are not aware of issues related to sustainability in software engineering, it is unlikely they will be wiling to teach it.
This calls for a change of their beliefs as sustainability is inevitable, and more and more present in a variety of disciplines. 
Such change of beliefs could be done via professional/staff development programs, seminars, workshops, etc \cite{ho2000conceptual}.\\
Our first recommendation is: \emph{Mobilize the unit or institution to raise awareness among educators regarding GSSE}.

As already argued in this paper, students in SE education should be made aware of the importance of sustainability. 
Sustainability should enter the curriculum not only as a standalone course, but as one of the learning objectives of the modules within a curricula.
This would position sustainability as one of the most valuable content within the curricula, placing it in the category of ``enduring understandings'' \cite{wiggins2005understanding}, which is knowledge that students remember even after they forget all other content.
However, the analysis of our survey shows that in SE the orientation is to convey sustainability topics through a standalone course.\\
Our second recommendation is: \emph{Include a specific learning objective targeting sustainability for each SE course}.

Educators perceive the development of teaching material as time-taking. 
However, this is a necessary activity that educators should accept through, for example, experimental learning \cite{kolb2014experiential}, and trial-and-error \cite{oleson2014teaching}. 
Educators will gain an understanding of what and how to teach, what are the issues that interest students and what matters the most to them. 
Such approach is appropriate given the current, early stage of GSSE.
As discussed earlier, educators answering the survey seem to favor a fact-based approach to teach GSSE. 
We believe that a discussion-based approach better suits GSSE as this offers the opportunity to generate teaching material by compiling notes, observations, and points of discussion that arise during the classes. 
Including students into the course design process \cite{kosnik2009improving} (i.e. invite students to contribute to decisions about course content and activities) can support the educators and at the same time raise the awareness and motivation of the students themselves, as they are creating material to be passed to their peers in the future \cite{prince2004does}.
The learner-centred and discussion-based approach, is more suitable to the nature of sustainability and it is the ways millennials learn best \cite{6227999}.\\
Our third recommendation is: \emph{to adopt a discussion-based teaching approach that involves students in the creation of the teaching material}.

\section{Threats to Validity}\label{sec:threats}
Construct Validity: the survey was conducted over the Internet so respondents might have misunderstood our questions. Nevertheless, to reduce ambiguity we reviewed the survey with colleagues, and we tried to ask questions which were very simple and straightforward. Our survey was a balanced mix of closed (without an option for the respondents' comments) and open questions. A closed questionnaire can improve participation rates, as it is easier to compile; conversely, open-ended questions improve the kind of feedback received.

%Conclusion Validity: as already stated in Section \ref{sec:method}, we did not carry out a quantitative statistical analysis of our data. This decision was made due to the limited sample size of our study. A quantitative analysis may overstate some of the relationships among our variables, leading to possible misinterpretation of the results \cite{Maxwell01072010}. As we do not perform statistical hypothesis testing, we are not concerned with conclusion validity threats.

Internal Validity: we did not carry out probabilistic sampling for the selection of the respondents. Our recruitment strategy could have incurred a possible selection bias (for example, a high probability of profile similarity among the respondents, such are respondents which are working in the context of GSSE). 
%It would be interesting to conduct a related survey with those who are not working in this area. We attempted to address this issue when we defined the protocol of the survey: we explicitly required the survey to be filled in by academics working in GSSE.

External Validity: the results are limited to those who have experience in the area of GSSE. 
%We make no claims that the results generalize to software engineering academics more broadly.

\section{Conclusion and Future Work}\label{sec:conclusion}
%\todo[author=Davide,inline,color=green]{Questa sezione va accorciata, le risposte a RQ possono essere succinte (una frase per RQ) e portate in Section VI.}
In recent years, great attention has been paid to sustainability issues. 
However, such issues seem to have only superficially interested software engineering higher education. Nowadays GSSE lends itself as an interesting area for institutions pledging for research-based education, as well as for teaching millenials, who display interest in sustainable green engineering~\cite{6227999}.
In this work, we presented a first attempt at showing the current status of GSSE in the universities curricula by surveying 33 academics in the community. 
In particular, we reported what are the topics of current interest, and how the related courses are organized. 
A first set of motivations behind the current presence (or lack) of GSSE in higher education curricula emerged from this study. 
As follows, we present our main findings according to our research questions:

\textit{RQ1}: The characterization of the sample shows that the responses come from researchers mostly located in Europe, and represent an active and young generation of the GSSE research community.  

\textit{RQ2}: The respondents feel the importance of tackling the social and environmental impact of software engineering; however, the technical dimension---already present in SE---is also deemed as important as the previous two.
GSSE topics have only superficially entered the curriculum, in fact the current lack of teaching in the respondents institution presents a duality: reinforcing the quality of the individual courses in terms of sustainability, without reducing the time allocated to the technical subjects.
This requires the resources to be well integrated within the curriculum, and be easily available to educators~\cite{Mann2010-zq}.

\textit{RQ3}: Although the respondents are involved in the GSSE community, the majority has not taught any related courses or modules. The lack of awareness was the main challenge in introducing GSSE in the university curricula. This supports our previous insight that sustainability is occasionally present in the curriculum. 
GSSE is taught either through one of more courses superficially focusing on GSSE, or through modules within existing courses. The courses, in which GSSE is addressed appear to be small in size (i.e., with no more than 20 students), and short in duration (i.e., maximum a semester long).The difficulties encountered when teaching these courses are related to the novelty of the field, which results in a lack of teaching material. Other hurdles are represented by the lack of awareness, and the multidisciplinary nature of the field.

\textit{RQ4}: The respondents suggest that the main topic to be taught is energy efficiency, and the main reasons for the lack of GSSE topics in higher education curricula are:
\begin{enumerate*}
\item lack of awareness,
\item lack of teaching material, 
\item high effort required, 
\item lack of technology and tool support.    
\end{enumerate*}
%\end{framed}

According to the latter, and based on current approaches from higher education research, we created a set of recommendations for educators.
We propose the following three recommendations when creating GSSE courses or curricula:
\begin{enumerate*}
\item \textit{mobilize  the  unit  or  institution to raise awareness among educators regarding GSSE};
\item \textit {include among the learning objectives of each SE course a specific one targeting sustainability};
\item \textit {adopt a discussion-based teaching approach that involves students in the creation of the teaching material}.
\end{enumerate*}
%\end{framed}

The above recommendations aim to address educators perspective on teaching about sustainable green engineering which is thus particularly significant as it will contribute towards developing pedagogies for sustainable green engineering that best suit millennial students needs.

The results from this first survey serve as a starting point for future work concerned with the integration and mainstreaming of green and sustainable software engineering education.
Our plans for further studies are different. 
Firstly, we are planning a follow--up set of in depth interviews with stakeholders in software engineering higher education to explore their prospectives on the topics identified in this paper, and their ideas about the feasibility of the suggested recommendations.
Secondly, we are going to extend this survey inviting researchers, practitioners, and academics (i.e. educators) from the entire SE community.
Finally, we aim to develop a set of concrete knowledge guidelines to help in designing GSSE course content, i.e. a concrete integration of GSSE topics in SE curricula.

\section*{Acknowledgments}
We would like to thank Trycia Bazinet (Carleton University, Ottawa, Canada) for the initial input about the paper.
\ifCLASSOPTIONcaptionsoff
  \newpage
\fi



%\bibliographystyle{IEEEtran}

\bibliographystyle{abbrv}
\bibliography{bibliography}
%\begin{biography}[{\includegraphics[width=1in,height=1.25in,clip,keepaspectratio]{mshell}}]{Michael Shell}


%\vfill


\end{document}